\documentclass[aps,pre,onecolumn]{revtex4}
\usepackage{epsfig,latexsym}

\begin{document}

\title{Distance traveled by random walkers before absorption in a
random medium}

\author{David S. Dean, Cl\'ement Sire,  and Julien Sopik}

\affiliation{
Laboratoire de Physique Th\'eorique (UMR 5152 du CNRS),
Universit\'e Paul Sabatier,\\
118, route de Narbonne, 31062 Toulouse Cedex 4,
France\\
E-mail: {\it dean@irsamc.ups-tlse.fr,
clement.sire@irsamc.ups-tlse.fr \& sopik@irsamc.ups-tlse.fr} }


\begin{abstract}
We consider the penetration length $l$ of random walkers diffusing
in a medium of perfect or imperfect absorbers of number  density
$\rho$. We solve this problem on a lattice and in the continuum in
all dimensions $D$, by means of a mean-field renormalization
group. For a homogeneous system in $D>2$, we find that $l\sim
\max\left(\xi,\rho^{-1/2}\right)$, where $\xi$ is the absorber
density correlation length. The cases of $D=1$ and $D=2$ are also
treated. In the presence of long-range correlations, we estimate
the temporal decay of the density of random walkers not yet
absorbed. These results are illustrated by exactly solvable toy
models, and extensive numerical simulations on directed
percolation, where the absorbers are the active sites. Finally, we
discuss the implications of our results for diffusion limited
aggregation (DLA), and we propose a more effective method to
measure $l$ in DLA clusters.
\end{abstract}

\maketitle

\section{Introduction}

The dynamics of random walkers diffusing in the presence of a
finite density of perfect absorbers is a rich problem which has
been widely discussed in the physical and mathematical literature
\cite{donsker}.  At very large times, the density of surviving
walkers does not decay exponentially as a simple mean-field
argument would predict, but rather behaves as
\begin{equation}
n(t)\sim\exp \left[-C_D\rho^{\frac{2}{D+2}}
t^{\frac{D}{D+2}}\right],
\label{don}
\end{equation}
where $\rho$ is the absorber density, and $C_D$ is a numerical
constant. The physical interpretation is that the process is
dominated by particles starting in very large absorber-free
regions (voids), of linear size $L$. In $D$ dimensions, these
regions have a probability of order
\begin{equation}
\exp(-\rho L^D),\label{volimp}
\end{equation}
for small $\rho$. In a void of size $L$, solving the diffusion
equation with absorbing conditions on its surface shows that the
density typically decays as $\exp(-t/L^2)$. A saddle-point
argument then leads to the result of Eq.~{(\ref{don})}, with the
relevant regions being of typical size $L\sim
(t/\rho)^{\frac{1}{D+2}}$, at time $t$.

Another important question is the determination of the penetration
or screening length $l$, which measures the average distance
between the starting point and the absorption point. In the limit
of a small density of uniformly distributed perfect absorbers of
radius $a$, and for $D>2$, a classic result \cite{muka} states
that
\begin{equation}
l\sim \frac{a^{1-D}}{\rho}.\label{faux}
\end{equation}
However, a simple heuristic argument casts doubts on the validity
of Eq.~(\ref{faux}). In $D$ dimensions, let us consider a
hypercubic box of linear size $L\sim\rho^{-1/D}$, which is the
typical distance between absorbers. We place one absorber of
radius $a$ in this box, and impose periodic boundary conditions,
which is equivalent to copying the box periodically. We now
release a random walker at time $t=0$. To estimate the typical
time when the walker will hit the absorber, we partition our box
in smaller boxes of size $a^D$. The random walker will be absorbed
with a finite probability once it has visited most of these small
cells, including in particular the one containing the absorber. We
define $N_D(t)$ as the number of different sites visited by a
discrete random walker after a time $t$. Hence, a fair estimate of
the absorbing time $t_{*}$ is given by
\begin{equation}
N_D(t_{*})\sim \left(\frac{L}{a}\right)^D\sim(\rho a^D)^{-1}.\label{eu1}
\end{equation}
During this time, the random walker has traveled a typical distance $l$
given by
\begin{equation}
l\sim \sqrt{\kappa t_{*}},\label{eu2}
\end{equation}
where $\kappa$ is the diffusion constant. We can use the classical
estimates for $N_D(t)$ \cite{itzy,feller}, which read
\begin{eqnarray}
N_1(t)&\sim &\frac{\sqrt{\kappa t}}{a},\qquad D=1,\label{eu31}\\
N_2(t)&\sim &\frac{\kappa t}{a^2\ln(\kappa t\, a^{-2})},\qquad
D=2,\label{eu32}\\ N_D(t)&\sim &\frac{\kappa t}{a^2},\qquad
D>2.\label{eu33}
\end{eqnarray}
Finally, combining
Eqs.~(\ref{eu1},\ref{eu2},\ref{eu31},\ref{eu32},\ref{eu33}), we
obtain the qualitative estimates
\begin{eqnarray}
l&\sim &\rho^{-1},\qquad  D=1,\label{eu41}\\
l&\sim &\sqrt{-\frac{\ln(\rho a^2)}{\rho}},\qquad  D=2,\label{eu42}\\
l&\sim &\frac{a^{1-D/2}}{\sqrt{\rho}},\qquad  D>2.\label{eu43}
\end{eqnarray}

In the present work, we will justify on more solid theoretical
grounds the results of Eqs.~(\ref{eu41},\ref{eu42},\ref{eu43}). In
section \ref{sec2}, we introduce a general formalism in order to
compute $l$ and the distribution of distances traveled before
absorption. In section \ref{sec3}, we define an exact
renormalization group for the Green's function on a lattice. This
recursion is solved using a mean-field (or cavity-like)
approximation and our result confirms the estimates of
Eqs.~(\ref{eu41},\ref{eu42},\ref{eu43}). In the limit $\rho\to 0$,
we also address the effect of imperfect absorbers, which becomes
relevant in $D>2$. In section \ref{sec4}, we extend this approach
to the continuum. We compute the distribution of the distances of
absorption and its first moments. The theory is found to be in
excellent agreement with numerical simulations in $D=2$ and $D=3$.
In section \ref{correla}, we apply our renormalization approach to
the case of a strongly correlated distribution of absorbers,
characterized by a power law decay of the absorber density
correlation function, $c(r)\sim r^{-\alpha}$, up to the
correlation length $\xi$. For $\alpha<2$ and $D>1$ (and for
$\alpha<1$, in $D=1$), the penetration length is found to be of
the same order as $\xi$, $l\sim \xi$. However, for $\alpha>2$,
correlation are weak enough, so that the uncorrelated result of
Eqs.~(\ref{eu41},\ref{eu42},\ref{eu43}) is recovered. This result
is illustrated by the exact solution of the problem in $D=1$, and
exactly solvable toy models in higher dimensions. In addition, we
test our ideas on the strongly correlated distribution of active
sites in directed percolation simulations performed at the
critical point, in $D=2$ and $D=3$. Random walkers absorbed by the
active sites are found to have a screening length $l\sim\xi$,
whereas the uncorrelated result is recovered above the critical
dimension $D_c=4$, when correlations become irrelevant. As a
conclusion of this section, we extend the result of
Eq.~{(\ref{don})} to the case of a strongly correlated density of
absorbers.  In section \ref{DLA}, we discuss the determination of
the screening length $l$ for diffusion limited aggregation (DLA)
clusters. We emphasize that the most common method to measure $l$
seems inappropriate and we propose an improved scheme. Finally, we
give some heuristic arguments inspired by the results of the
previous sections, leading to the estimate $l\sim R$, for DLA
clusters of gyration radius $R$.

\section{General Backward Fokker-Planck Formalism}\label{sec2}

In a $D$-dimensional space, we consider a Brownian particle with
diffusion constant $\kappa$. Let $V({\bf x})$ be a positive
killing field such that if the Brownian particle is a the position
${\bf x}$, then in the next time interval $dt$, it is killed with
probability $V({\bf x}) \,dt$. If the particle is not killed, it
just keeps on diffusing. We define $P({\bf x},{\bf y})$  as the
probability density that, starting from ${\bf x}$, that the
particle's last resting place, {\em i.e.} where it is killed or
absorbed, is at ${\bf y}$. The quantity $P({\bf x}, {\bf y})$ can
be calculated by standard backward Fokker-Planck techniques. We
consider what happens in the first time step $dt$. One
possibility, occurring with probability $V({\bf x}) \,dt$, is that
the particle is killed where it starts. The other possibility is
that it is not killed, this with probability $1- V({\bf x})\, dt$,
and so can then make a Brownian jump $d{\bf B}$ having a component
in the spatial direction $\mu$, $dB_\mu$, obeying $\langle
dB_{\mu}dB_{\mu'}\rangle = 2\kappa \delta_{\mu\mu'}dt$. Putting
this together gives
\begin{equation}
P({\bf x},{\bf y}) = \delta({\bf x}-{\bf y})V({\bf x}) dt +
\left(1 - V({\bf x}) dt\right)\langle P({\bf x} + d{\bf B},{\bf
y})\rangle.
\end{equation}
Expanding to order $dt$, and taking the expectation value over
$d{\bf B}$, we obtain
\begin{equation}
-\kappa\nabla_{\bf x}^2 P({\bf x},{\bf y}) + V({\bf x})P({\bf x},
{\bf y}) = \delta({\bf x}-{\bf y})V({\bf y})\label{eqP}.
\end{equation}
The solution to Eq.~(\ref{eqP}) is given by
\begin{equation}
P({\bf x},{\bf y}) = G({\bf x},{\bf y}) V({\bf y}),
\end{equation}
where $G$ is the Green's function obeying
\begin{equation}
-\kappa\nabla_{\bf x}^2 G({\bf x},{\bf y}) + V({\bf x})
G({\bf x},{\bf y}) = \delta({\bf x}-{\bf y})\label{eqG}.
\end{equation}
Note that the derivation of a probability density rather than a
probability always requires a bit of care and our above derivation
can be made more rigorous by defining an interval $A$ around the
point ${\bf y}$ then computing the probability that the particle
is killed in $A$ then taking the limit $A\to 0$. One can check
that $P$ is normalized as follows. Clearly the operator acting on
$G$ in its defining equation Eq.~(\ref{eqG})  is self-adjoint,
which means that $G({\bf x},{\bf y})= G({\bf y},{\bf x})$.
Integrating Eq.~(\ref{eqG}) over all ${\bf x}$, and if the
potential $V$ is sufficiently strong, the first term of
Eq.~(\ref{eqG}) will give an irrelevant surface term. We thus
obtain
\begin{equation}
\int d{\bf x} \,V({\bf x})G({\bf x},{\bf y}) = 1.
\end{equation}
However $G$ is symmetric which ensures that
\begin{equation}
\int d{\bf y} \,V({\bf y})G({\bf x},{\bf y}) = \int d{\bf y}\,
P({\bf x},{\bf y}) = 1,
\end{equation}
and which demonstrates the correct normalization of $P$. To
further simplify this problem and reduce everything to the study
of the Green's function, we write $H({\bf x},{\bf y}) = -\kappa
\nabla_{\bf x}^2 \delta({\bf x}-{\bf y})$ and consider $V$ as an
operator, $V({\bf x},{\bf y}) = \delta({\bf x}-{\bf y})V({\bf
y})$. Using operator notation, $P$ is given by
\begin{equation}
P = [H+V]^{-1} V =  [H+V]^{-1} [H +  V -H] = I - [H+V]^{-1} H,
\end{equation}
which reads, in coordinate notation,
\begin{equation}
P({\bf x},{\bf y}) = \delta({\bf x}-{\bf y}) + \kappa \nabla^2_{\bf y}
G({\bf x},{\bf y}).
\end{equation}
Now defining the disorder-averaged values of $P$ and $G$ by $p$ and
$g$ respectively, we can write the averaged form of the above equation
for a translational invariant distribution of absorbers
\begin{equation}
p({\bf y}) = \delta({\bf y}) + \kappa \nabla^2_{\bf y} g({\bf y}).
\label{eqpy}
\end{equation}
If the disorder is isotropic, we will have $p({\bf y}) = p(y)$,
where $y = |{\bf y}|$, and likewise for $g$. The disordered
averaged moments of the  distance traveled before the particle is
killed are more suitably obtained from $\tilde p$, the Fourier
transform of $p$ defined as
\begin{equation}
{\tilde p}({\bf k}) = \int d{\bf x} \exp(-i{\bf k}\cdot {\bf x})
p({\bf x}).
\end{equation}

If we write the Fourier
transform of $g$, $\tilde g$ as
\begin{equation}
{\tilde g}(k) = {1\over \kappa k^2+ \kappa\Sigma(k)},
\end{equation}
where $\kappa\Sigma(k)$ is given by the one particle irreducible
diagrams, we then obtain
\begin{equation}
{\tilde p}(k) = {\Sigma(k)\over k^2+ \Sigma(k)}.\label{eqfg}
\end{equation}
Now, for small $k$ and to leading order, we expect that
\begin{equation}
\Sigma(k) \approx m^2 + {\delta\kappa\over \kappa} k^2 + O(k^4),
\end{equation}
where $m$ is the inverse effective screening length of the
averaged Green's function and $\delta\kappa$ is the
renormalization of the  diffusion constant. The  resulting small
$k$ expansion for $\tilde p$ is
\begin{equation}
{\tilde p}(k) \approx 1 - {k^2\over m^2}+O(k^4),
\end{equation}
so if ${\bf Y}$ is the position where a particle released at the
origin is observed, then the disorder averaged second moment of
${\bf Y}$ is simply given by
\begin{equation}
\overline{\langle {\bf Y}^2 \rangle} = {2D\over m^2}.
\end{equation}
The characteristic distance from the starting position, $l$, at
which the particle gets absorbed is therefore the same as the screening
length for the Green's function. We thus define
\begin{equation}
l=\sqrt{\overline{\langle {\bf Y}^2 \rangle}} = {\sqrt{2D}\over
m}\sim m^{-1}.
\end{equation}

On a discrete lattice, the same arguments apply and we find that
\begin{equation}
p({\bf y}) = \delta_{{\bf y},{\bf 0}} + \nabla^2 g({\bf y}),
\end{equation}
where $\nabla^2$ denotes the lattice Laplacian. The discrete
Fourier transform of $p$ is defined as
\begin{equation}
{\tilde p}({\bf k}) = \sum_{\bf x} \exp(-i{\bf k}\cdot{\bf x}) p({\bf x}),
\end{equation}
and is given by
\begin{equation}
{\tilde p}({\bf k}) = 1 + \left[2\sum_\mu \cos(k_\mu)  - 2D\right]
{\tilde g}({\bf k}),
\end{equation}
on a $D$-dimensional cubic lattice of lattice spacing $a=1$. Now,
if we write
\begin{equation}
{\tilde g}({\bf k}) = {1\over   2D- 2\sum_\mu \cos(k_\mu)  +
\Sigma({\bf k})},
\end{equation}
we find the lattice result, analogous to Eq.~(\ref{eqfg})
\begin{equation}
{\tilde p}({\bf k}) = {\Sigma({\bf k})\over 2D- {2}\sum_\mu
\cos(k_\mu)  + \Sigma({\bf k})}.
\end{equation}
The leading order behavior of $\Sigma$ must be of
the form
\begin{equation}
\Sigma({\bf k}) \approx m^2 + \delta\kappa\, k^2+O(k^4),
\end{equation}
thus yielding the lattice result
\begin{equation}
l=\sqrt{\overline{\langle {\bf Y}^2 \rangle}} = {\sqrt{2D}\over
m},
\end{equation}
which takes the same form as in the continuum. Interestingly, we
find that $l$ is not affected explicitly by the renormalization of
the diffusion constant $\delta\kappa$.

\section{Mean-field renormalization group calculation on a lattice}\label{sec3}

In this section, we will consider a model on the lattice $Z^D$
(hence the lattice constant is $a=1$) with a potential $V_N({\bf
x})$ given by
\begin{equation}
V_N({\bf x}) = \lambda \sum_{i=1}^N \delta_{{\bf x},{\bf a}_i}=
\sum_{i=1}^N U_i({\bf x}).
\end{equation}
where the ${\bf a}_i$'s are absorbing sites, with strength
$\lambda$, which are uniformly and independently (in this first
instance) distributed among all lattice sites. We denote by ${\cal
V}$ the total number of lattice sites.

We now estimate the renormalization of the Green's function $G_N$
for a system with $N$ absorbing sites by the addition of an
$N+1$-th absorbing site. A similar method has been introduced in
\cite{dean} to calculate the effective diffusion constant of a
tracer particle in a medium composed of randomly placed
scatterers. We denote by ${\bf a}_{N+1}$ the position of the newly
added absorber. In operator notation, and introducing the discrete
Laplacian $H=-\kappa\nabla^2$, the discrete version of
Eq.~(\ref{eqG}) for $G_N$ and $G_{N+1}$ leads to,
\begin{eqnarray}
(H+V_N)G_N&=&I,\\
(H+V_N+U_{N+1})G_{N+1}&=&I.
\end{eqnarray}
Eliminating $H+V_N=G_N^{-1}$, we find
\begin{equation}
G_{N}^{-1} G_{N+1} + U_{N+1} G_{N+1} = I.
\end{equation}
After multiplying by $G_N$ and using the explicit form of
$U_{N+1}$, we obtain in coordinates notation,
\begin{equation}
G_{N+1}({\bf x},{\bf y}) = G_{N}({\bf x},{\bf y}) -\lambda
G_{N}({\bf x},{\bf a}_{N+1})G_{N+1}({\bf a}_{N+1},{\bf
y}).\label{recnew}
\end{equation}
If we set ${\bf x}={\bf a}_{N+1}$, we obtain a closed form
expression for $G_{N+1}({\bf a}_{N+1},{\bf y})$, which can be
substituted into Eq.~(\ref{recnew}). This leads to the following
{\it exact} recurrence:
\begin{equation}
G_{N+1}({\bf x},{\bf y}) = G_{N}({\bf x},{\bf y}) -{\lambda
G_{N}({\bf x},{\bf a}_{N+1})G_{N}({\bf a}_{N+1},{\bf y})\over 1 +
\lambda G_{N}({\bf a}_{N+1},{\bf a}_{N+1})}.\label{eqg1l}
\end{equation}

Let us first consider the limit where the absorbing sites kill the
particle on contact with probability $1$. This is achieved by
taking the limit $\lambda\to +\infty$ and yields
\begin{equation}
G_{N+1}({\bf x},{\bf y}) = G_{N}({\bf x},{\bf y}) -{G_{N}({\bf
x},{\bf a}_{N+1})G_{N}({\bf a}_{N+1},{\bf y})\over G_{N}({\bf
a}_{N+1},{\bf a}_{N+1})}, \label{eqg1}
\end{equation}
which describes the exact renormalization of the Green's function
due to the addition of a perfect absorber.

If the perfect absorbers are independently distributed, the
position ${\bf a}_{N+1}$ is completely uncorrelated with the ${\bf
a}_i$'s, for $i=1,\cdots, N$. Averaging Eq.~(\ref{eqg1}) over the
position ${\bf a}_{N+1}$, but ignoring the correlation between the
numerator and the denominator in the second term (a
mean-field-like approximation), we obtain
\begin{equation}
G_{N+1}({\bf x},{\bf y}) =   G_{N}({\bf x},{\bf y}) - {\sum_{\bf
a}G_{N}({\bf x},{\bf a})G_{N}({\bf a},{\bf y})\over \sum_{\bf
a}G_{N}({\bf a},{\bf a})}.
\end{equation}
We now perform the average over the remaining particle positions.
Using the statistical translational invariance of the system, we
find
\begin{equation}
g_{N+1}({\bf y}) =   g_{N}({\bf y}) - {1\over {\cal V}}{\sum_{\bf
x}g_{N}({\bf x})  g_N({\bf y}-{\bf x}) \over g_N({\bf 0})}.
\end{equation}
Taking the Fourier transform of this equation yields
\begin{equation}
{\tilde g}_{N+1}({\bf k}) =  \tilde{g}_{N}({\bf k}) -
{1\over {\cal V}}{\tilde{g}^2_{N}({\bf k}) \over
g_N({\bf 0})}.
\end{equation}
Defining the density $\rho = N/{\cal V}$, we obtain a differential
equation for the evolution of ${\tilde g}(\rho,{\bf k})$ in this
approximation
\begin{equation}
{\partial{\tilde g}\over \partial \rho} (\rho,{\bf k}) =
-{\tilde{g}^2(\rho,{\bf k})\over g(\rho,{\bf 0})},\label{eqdif}
\end{equation}
where $g(\rho,{\bf 0})$ is self-consistently given by
\begin{equation}
g(\rho,{\bf 0}) = \int_{-\pi}^{\pi} {d{\bf k}\over (2\pi)^{D}}
{\tilde g}(\rho,{\bf k}).\label{eqdifsc}
\end{equation}
The solution to Eq.~(\ref{eqdif},\ref{eqdifsc}) with the correct
boundary conditions is
\begin{equation}
\tilde{g}(\rho,{\bf k})= {\kappa^{-1}\over   2D- 2\sum_\mu
\cos(k_\mu) + s(\rho)},\label{propag}
\end{equation}
where $s(\rho)$ obeys
\begin{equation}
{ds\over d\rho}=   \left[\int_{-\pi}^{\pi} {d{\bf k}\over
(2\pi)^{D}} {1\over  2D- 2\sum_\mu \cos(k_\mu)  +
s(\rho)}\right]^{-1}.\label{des}
\end{equation}
As expected, the defining equation for $s(\rho)$ does not depend
on $\kappa$ for perfect absorbers.

Now using the fact that $s(0)=0$, we can integrate Eq.~(\ref{des})
to obtain
\begin{equation}
\int_{-\pi}^{\pi} {d{\bf k}\over (2\pi)^{D}} \ln\left( 1 +
{s(\rho)\over 2D- 2\sum_\mu \cos(k_\mu)} \right) = \rho.
\end{equation}
For small $s$, we can use standard results for the integral in
Eq.~(\ref{des}) \cite{itzy}. The results depend on the
dimensionality and the lattice structure.

\begin{itemize}
\item {\bf $D=1$}: Substituting in the small $s$ behavior of
the  integral in Eq.~(\ref{des}) \cite{itzy}, we find
\begin{equation}
{ds\over d\rho}\approx 2\sqrt{s}.
\end{equation}
For small $\rho$, integrating this gives,
\begin{equation}
s=m^2={\rho^2},
\end{equation}
which then leads to
\begin{equation}
l=\frac{\sqrt{2}}{m}=\frac{\sqrt{2}}{\rho}.\label{resul1ds}
\end{equation}
This is clearly the correct scaling in $\rho$, which will be
recovered when solving exactly the one dimensional case (see
section \ref{exact1d}). Formally, the result $l\sim \rho^{-1/D}$
holds for any dimension $1\leq D <2$. The physical interpretation
is clear: the screening length is simply proportional to the mean
distance between absorbers, a result which will only remain true
in the absence of strong correlation between them.

\item{\bf $D=2$}: Here we find \cite{itzy}
\begin{equation}
{ds\over d\rho}\approx -{4\pi\over \ln(s)},
\end{equation}
and integrating this for small $\rho$, we obtain
\begin{equation}
s=m^2\approx -{4\pi \rho\over \ln(\rho)}.
\end{equation}
Hence, the screening length is
\begin{equation}
l=\frac{2}{m} = \sqrt{-\frac{\ln(\rho)}{\pi
\rho}}.\label{resul2ds}
\end{equation}

\item{\bf $D>2$}: Here, the leading order is analytic in $\rho$
\cite{itzy}, and we find
\begin{equation}
{ds\over d\rho}\approx {1\over g(0,{\bf 0})},
\end{equation}
giving
\begin{equation}
l=\frac{\sqrt{2D}}{m}=\sqrt{\frac{2D g(0,{\bf
0})}{\rho}}.\label{resul3ds}
\end{equation}
The term $g(0,{\bf 0})$ depends explicitly on the dimension and
the lattice structure. For example, for the three dimensional
cubic lattice, one has $g(0,{\bf 0})\approx 0.25...$ \cite{itzy}.
\end{itemize}

Note that we expect these results to be strongly modified if the
absorbers positions are spatially correlated, a problem which will be
addressed in section \ref{correla}.

Let us comment on the disagreement between the present results and
the one of Ref.~\cite{muka}, presented in Eq.~(\ref{faux}). In
\cite{muka}, the author first treats the effect of a single
absorber on the free Green's function. He then assumes that the
total correction to $\Sigma({\bf k})$ is simply proportional to
the number of absorbers. This statement is in fact incorrect, as
absorbers far away from the introduced random walker should have a
negligible contribution. Indeed, the walker should be absorbed
well before being able to visit the regions where they stand. This
is actually the effect of screening, that our renormalization
approach effectively captures. In addition, Eq.~(18) in
Ref.~\cite{muka}, which leads to the final result of
Eq.~(\ref{faux}), does not make any sense in the small $\rho$
limit, and it seems that the opposite non physical limit was in
fact taken.

Finally, we can extend our formalism to the case of imperfect
absorbers, corresponding to a finite value of $\lambda$. The
penetration length $l$ is now expected to depend explicitly on the
diffusion constant $\kappa$. Using Eq.~(\ref{eqg1l}), we find that
the Green's function still takes the form of Eq.~(\ref{propag}),
but with $s(\rho)$ now satisfying
\begin{equation}
{ds\over d\rho}=   \left[\delta+\int_{-\pi}^{\pi} {d{\bf k}\over
(2\pi)^{D}} {1\over  2D- 2\sum_\mu \cos(k_\mu)  +
s(\rho)}\right]^{-1},\label{desnewl}
\end{equation}
with
\begin{equation}
\delta=\frac{\kappa}{\lambda}>0.
\end{equation}
In one and two dimensions a random walk is recurrent and visits
any position a large number of times. If the density of absorbers
is small enough, a particle will  visit the first absorber
encountered a large number of times before making an excursion
sufficiently far away from this first absorber and visiting a
region occupied by a different absorber. This means that most
particles are absorbed by the first absorber encountered and thus
the effect of a finite $\lambda$ should become irrelevant in
$D\leq 2$, when $\rho\to 0$.

In $D=1$, we find the explicit result
\begin{equation}
\sqrt{s}+\delta s=\rho,
\end{equation}
which leads to
\begin{equation}
l=\frac{2\sqrt{2}\delta}{\sqrt{1+4\delta\rho}-1}.
\end{equation}
Hence, for $\rho\ll\delta^{-1}$, we recover the result of
Eq.~(\ref{resul1ds}).

In $D=2$, Eq.~(\ref{desnewl}) leads to
\begin{equation}
 s\left[\delta -{ \ln(s)\over 4\pi}\right]=\rho,
\end{equation}
or
\begin{equation}
l=\sqrt{-\frac{\ln(\rho)}{\pi \rho}+\frac{4\delta}{\rho}}.
\end{equation}
For $\rho\ll\exp(-4\pi\delta)$, we recover the result of
Eq.~(\ref{resul2ds}).

Finally for $D>2$, considering imperfect absorbers deeply affects
the result of Eq.~(\ref{resul3ds}). Indeed, we obtain
\begin{equation}
l=\sqrt{\frac{2D (g(0,{\bf 0})+\delta)}{\rho}}.
\end{equation}
Note that in all dimensions, we find that the penetration length
is an increasing function of $\delta=\frac{\kappa}{\lambda}$, as
physically expected.

\section{The screening length in the continuum}\label{sec4}
\subsection{Delta function absorbers in one dimension}\label{exact1d}
We consider a system where the particle performs continuous
Brownian motion in one dimension. It makes perfect sense to take
an absorbing potential of the form
\begin{equation}
V(x) = \lambda\sum_{i=1}^N \delta(x-a_i),
\end{equation}
corresponding to point-like absorbers. Here, we consider the case
where the random walker is absorbed with probability one at each
absorbing site, that is to say the limit $\lambda\to +\infty$.
Without loss of generality, we set $\kappa=1$, since the final
expression of $l$ for perfect absorbers cannot depend on the value
of $\kappa$, whatever the spatial dimension. As in the previous
section, we apply the cavity approach to calculate the
renormalization of the Green's function by the addition of an
extra absorber at $a_{N+1}$. We apply again the method of section
\ref{sec3}, which leads to the recurrence equation
\begin{equation}
{d{ \tilde g}\over d\rho} = - {{ \tilde g}^2(\rho,k)\over
g(\rho,0)},\label{dif1d}
\end{equation}
where the continuous Fourier transform of $g$ is defined as
\begin{equation}
 g(\rho,x) = {1\over 2\pi} \int_{-\infty}^{\infty} dk\, \exp(ikx)
{\tilde g}(\rho,k).
\end{equation}
The solution of  Eq.~(\ref{dif1d}) is given by
\begin{equation}
{\tilde g} = {1\over k^2 + s(\rho)},
\end{equation}
where $s$ satisfies
\begin{equation}
{ds\over d\rho} = 2\sqrt{s},
\end{equation}
which leads to $s=m^2 = \rho^2$. Hence, we recover the lattice
result
\begin{equation}
l=\frac{\sqrt{2}}{m}=\frac{\sqrt{2}}{\rho}.
\end{equation}
We also obtain the explicit form
\begin{equation}
{\tilde p}(k) = {\rho^2\over k^2 + \rho^2},
\end{equation}
which gives the distribution of $Y$, the displacement from the
starting position to the point of absorption, to be
\begin{equation}
p(y) = {\rho\over 2} \exp(-\rho|y|).
\end{equation}

In fact, in this one dimensional case, the distribution of $Y$ can
be computed exactly. A random walker starting from $x=0$, we
denote by $b$ the closest absorbing site to the right and by $-a$
the closest absorbing site to the left. Standard results on
Brownian motion \cite{feller} tell us that the probability of
hitting $b$ before $-a$ is
\begin{equation}
p_b = {a\over a+b}.
\end{equation}
This means that the probability density function for $Y$ before
averaging over the disorder is simply
\begin{equation}
P(y) = {a\over a+b}\delta(y-b) + {b\over a+b}\delta(y+a).
\end{equation}
The disordered averaged density $p(y)$ is now given by
\begin{equation}
p(y) = \langle P(y)\rangle_{a,b}\,,
\end{equation}
where the angled brackets denote the average over the absorber
positions $a$ and $b$. If the absorbers are placed as a Poisson
point process with rate $\rho$, then the probability density
function of $a$ and $b$ is Poissonian
\begin{equation}
w(x) = \rho \exp(-\rho x).
\end{equation}
This yields
\begin{equation}
p(y) = \rho\exp(-\rho|y|)\int_0^\infty dx\, {x\exp(- x) \over x +
\rho |y|}=\rho\int_{\rho|y|}^\infty dx \left(1-{\rho|y|\over
x}\right)\exp(- x),
 \label{eqpo}
\end{equation}
showing that in this case the relevant length scale in indeed
$l\sim \rho^{-1}$. However, we note that the probability density
function given by the mean-field renormalization method is not
exact in one dimension.

A generalization of this one dimensional model can be constructed
as follows. We take the distribution of lengths $L$ between the
absorbers to be given by $q(L)$. The average interval length is
simply related to the density by
\begin{equation}
\rho^{-1}=\langle L\rangle = \int_0^\infty dL\, Lq(L).
\end{equation}
The probability that the particle starts within an interval of
length $L$ has the probability distribution function
\begin{equation}
Q(L) = {L\over \langle L\rangle}q(L).\label{qq}
\end{equation}
Given that one is in this interval, the position within it is
uniformly distributed. This means that we can write $b = L(1-U)$
and $a= LU$, where $U$ is uniformly distributed over $[0,1]$. We
thus have
\begin{equation}
P(y) =    U\delta(y-L(1-U)) + (1-U)\delta(y+LU).
\end{equation}
Now performing the disorder average and using the symmetry of the
problem, we find
\begin{eqnarray}
p(y)&=& \int_0^1 du \int_0^\infty dL\, u Q(L)  \,
\delta(|y| - (1-u)L),\nonumber \\
&=&  \int_0^1 du \ {u\over 1-u}Q\left({|y|\over 1-u}\right).
\end{eqnarray}
Using the explicit form for $Q$ in Eq.~(\ref{qq}), we obtain
\begin{equation}
p(y) = {\int_{|y|}^\infty dx\, \left(1-{|y|\over
x}\right)q(x)\over \int_0^\infty dx\ x q(x) }.
\end{equation}
As a check,  we set $q(x) = \rho\exp(-\rho x)$ and recover the
result Eq.~(\ref{eqpo}), obtained for the memoryless distribution
of Poissonian absorbers.

The above method of calculation also allows for a straightforward
computation of the moments of $|Y|$ which can be conveniently
expressed in terms of the moments of the distribution $q(L)$. We
find
\begin{equation}
\langle |Y|^n\rangle = {2\over (n+1)(n+2)}{\langle L^{n+1}\rangle
\over \langle L\rangle}.\label{lexactmom1}
\end{equation}
In particular, we have
\begin{equation}
l^2=\langle |Y|^2\rangle =\frac{1}{6}{\langle L^{3}\rangle \over
\langle L\rangle},\label{lexact}
\end{equation}
which gives
\begin{equation}
l=\rho^{-1}=\langle L\rangle,
\end{equation}
for uniformly distributed absorbers associated to a Poissonian
distribution of intervals.

\subsection{Absorbing spheres in $D\geq 2$}

In this section, we compute the behavior of the distance traveled
before absorption in two dimensions and above. In this case, we
take the absorbers to be spheres of radius $a$. As before, we
restrict ourselves to the limit where the density of absorbers
$\rho = N/{\cal V}$ is small. As in the case of the lattice
system, we denote by $G_N$ the Green's function in the presence of
$N$ absorbers and by $G_{N+1}$ the Green's function obtained when
an extra absorber is placed at the point ${\bf a}_{N+1}$ which is
uniformly distributed in the volume ${\cal V}$, independently of
the other absorbers.  We denote by $U_{N+1}({\bf x})$ the
potential due to the $N+1$-th absorber which takes the form
\begin{equation}
U_{N+1}({\bf x}) = \lambda\,\delta(a - |{\bf x} -\bf {a}_{N+1}|),
\end{equation}
where $\lambda$ has the same dimension as $\kappa$ divided by a
distance. $U_{N+1}$ is absorbing with strength $\lambda$ on the
{\it surface} of the sphere of radius $a$ centered at the point
${\bf a}_{N+1}$. In the limit $\lambda \to +\infty$, the absorber
is a perfect absorber. Combining Eq.~(\ref{eqG}) written for $G_N$
and $G_{N+1}$, we can find the equation for $G_{N+1}$ in operator
notation as
\begin{equation}
G_{N}^{-1} G_{N+1} + U_{N+1} G_{N+1} = I,
\end{equation}
which can be rewritten as
\begin{equation}
G_{N+1}({\bf x},{\bf y}) = G_{N}({\bf x},{\bf y}) - \int d{\bf
z}\, G_{N}({\bf x},{\bf z})U_{N+1}({\bf z}) G_{N+1}({\bf z},{\bf
y}).
\end{equation}
The corresponding equation on the lattice was easy to solve as the
potential was given by a delta function. However, here we must resort
to a further approximation: we assume that we can set $ G_{N}({\bf
x},{\bf z}) \approx G_{N}({\bf x},{\bf a}_{N+1})$ and $ G_{N}({\bf
z},{\bf y}) \approx G_{N}({\bf a}_{N+1},{\bf y})$ in the above
integral over the surface of a sphere centered at ${\bf a}_{N+1}$. We
obtain
\begin{equation}
G_{N+1}({\bf x},{\bf y}) = G_{N}({\bf x},{\bf y}) -
{\lambda\sigma(D) a^{D-1}G_{N}({\bf x},{\bf a}_{N+1}) G_{N}({\bf
a}_{N+1},{\bf y})\over 1 + \int d{\bf z} \, U_{N+1}({\bf z})
G_{N}({\bf z},{\bf a}_{N+1})},\label{approx}
\end{equation}
where $\sigma(D)$ is the area of a sphere of unit radius in $D$
dimensions. We cannot make the same approximation for the integral
in the denominator as the resulting term would be proportional to
$ G_{N}({\bf a}_{N+1},{\bf a}_{N+1})$ which is finite on a
lattice, but diverges in the continuous case for $D\geq 2$. The
finite radius of the absorbing spheres regularizes the result.
Now, in the limit of large $\lambda$, we obtain the expression for
the renormalization group flow
\begin{equation}
G_{N+1}({\bf x},{\bf y}) = G_{N}({\bf x},{\bf y}) - {\sigma(D)
a^{D-1}G_{N}({\bf x},{\bf a}_{N+1}) G_{N}({\bf a}_{N+1},{\bf
y})\over \int d{\bf z}\, I({\bf z},{\bf a}_{N+1}) G_{N}({\bf
z},{\bf a}_{N+1})},
\end{equation}
where
\begin{equation}
I({\bf x},{\bf a}_{N+1}) = \delta(a - |{\bf x} -\bf {a}_{N+1}|).
\end{equation}
Following the same line of arguments as in the lattice case, we
find that the disorder averaged Green's function obeys
\begin{equation}
{d{ \tilde g}\over d\rho} = - {{ \tilde g}^2(\rho,k)\over g(a)},
\end{equation}
where we have used the  spherical symmetry of the disorder
averaged Green's function. For $\kappa=1$, the solution to this
equation is
\begin{equation}
{ \tilde g} = {1\over k^2 + s},
\end{equation}
where $s$ obeys
\begin{equation}
{d{ s}\over d\rho} =  { 1\over g(a)}\label{eqmas},
\end{equation}
and where $g(a)$ has to be computed self-consistently.
\begin{itemize}
\item{$D=2$}: In two dimensions, we have
\begin{equation}
g(r) = {1\over 2\pi}K_0(\sqrt{s} r),
\end{equation}
where $K_0(x)$ is the Bessel function of the second kind of order 0
\cite{abrom}. We thus have
\begin{equation}
{d{ s}\over d\rho} = { 2\pi \over K_0(\sqrt{s} a)}.
\end{equation}
Assuming that $s$ is sufficiently small, {\em i.e.} for
sufficiently small $\rho$, we can use the small argument
asymptotic form of $K_0$ \cite{abrom}, $K_0(u) \sim -\ln(u)$, to
obtain
\begin{equation}
{d{ s}\over d\rho} = { 4\pi\over -\ln(sa^2)},
\end{equation}
where a cut-off of order $a^2$ naturally arises. This equation can
be integrated up to the leading order, yielding
\begin{equation}
m^2 = s \approx -{4\pi \rho\over \ln(\rho a^2)}. \label{eqm3d}
\end{equation}
We thus find the same functional form as in the lattice case.
After some algebra, we find the following results for the lowest
order moments:
\begin{equation}
\langle |{\bf Y}|\rangle = {1\over 4} \sqrt{-\pi{\ln(\rho
a^2)\over \rho}},\label{y12d}
\end{equation}
and
\begin{equation}
l=\sqrt{\langle{\bf Y}^2\rangle} = \sqrt{-{\ln(\rho a^2)\over
\pi\rho}}.\label{y22d}
\end{equation}
The above calculation actually gives the full distribution for
${\bf Y}$. If $r =|{\bf Y}|$, then the probability density
function for $R$ is given by
\begin{equation}
p(r) = m^2rK_0(mr),
\end{equation}
where $m$ is given by Eq.~(\ref{eqm3d}). In the case of imperfect
absorbers (finite $\lambda$), the present results are not affected
provided that $\rho$ is small enough ($-\delta/\ln(\rho)\ll 1$).

In Fig.~\ref{fig2d}, we plot the two first moments of ${\bf Y}$,
as found from numerical simulations. We find a perfect agreement
with Eqs.~(\ref{y12d},\ref{y22d}), indicating that the
theoretically predicted constant prefactors of the $\rho$
dependence are in fact probably exact in the small density limit.
Numerically, in order to access to the low density regime, we use
the following algorithm which has been introduced in the context
of DLA \cite{meakin}. Before performing the next move of our
Brownian walker, we look for the nearest absorber (by inspection
of a grid keeping track of the coarse-grained absorber density),
say found at the distance $d$. We then deposit the random walker
anywhere on the circle of radius $d-a$ centered at its current
position. Indeed, the first position where the walker would cross
the perimeter is uniformly distributed on the circle. The random
walker is absorbed when the new distance to its nearest absorber
is $d\leq a(1+\varepsilon)$, where $\varepsilon$ is small enough
(typically, we took $\varepsilon\leq 10^{-3}$, and $a=1$). Note
that if we were to actually simulate the random walker trajectory
before it reaches the perimeter, the program would take a running
time larger by a factor $d^2/dt$, where $dt$ is the time
increment. When accessing to density of order $\rho\sim 10^{-6}$,
and using a time step of order $dt\sim 10^{-3}$, this factor is of
order $10^9$, which gives an idea of the huge gain achieved by
using this algorithm. All the simulations performed in this paper
use variants of this algorithm.

\begin{figure}
\psfig{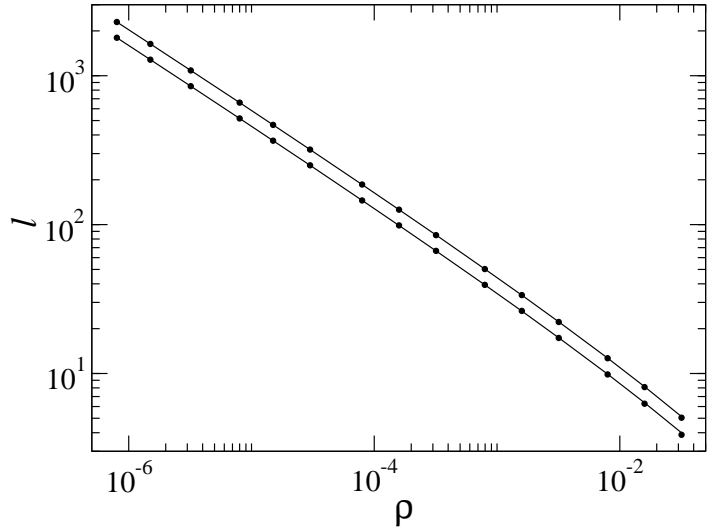}
\caption{\label{fig2d} Plot of $\langle |{\bf Y}|\rangle$ (bottom
full circles) and $l=\sqrt{\langle{\bf Y}^2\rangle}$ (top full
circles) obtained from numerical simulations in $D=2$. Each point
is obtained from the average over at least $10^6$ random walker
trajectories, and we used several samples totalizing at least
$10^6$ uniformly distributed absorbers. Error bars are much
smaller than the size of the circles. We compare these data to the
functional forms of Eqs.~(\ref{y12d},\ref{y22d}) (full lines),
with the cut-off $a^2$ being the sole fitting parameter. For
spheres of radius $a=1$, we find that both curves are well fitted
with an effective value of $a^2\approx 2.2$.}
\end{figure}

\item{$D=3$}: In this case, we find
\begin{equation}
g(r) = {1\over 4\pi r}\exp(-\sqrt{s} r).
\end{equation}
For small $s$, Eq.~(\ref{eqmas}) yields
\begin{equation}
m^2 = s\approx {4\pi a}\rho,
\end{equation}
and we again see that the functional dependence is the same as for
the lattice case. The  lowest order moments are given by
\begin{equation}
\langle |{\bf Y}|\rangle = \sqrt{1\over \pi \rho a},\label{y13d}
\end{equation}
and
\begin{equation}
l=\sqrt{\langle{\bf Y}^2\rangle} = \sqrt{3\over 2\pi \rho a}.
\label{y23d}
\end{equation}
For imperfect absorbers, the result now depends on the diffusion
constant
\begin{equation}
l= \sqrt{{3\over 2\pi \rho a}\left(1+\frac{\kappa}{\lambda
a}\right)}. \label{y23dk}
\end{equation}
The radial distribution function for ${\bf Y}$ takes the explicit
form
\begin{equation}
p(r) = m^2 r \exp(-mr).\label{prth}
\end{equation}
In Fig.~\ref{fig3d}, we plot the two first moments of ${\bf Y}$,
as found from numerical simulations. We find a perfect agreement
with Eqs.~(\ref{y13d},\ref{y23d}), indicating that these
expressions are again probably exact in the small density limit.
\begin{figure}
\psfig{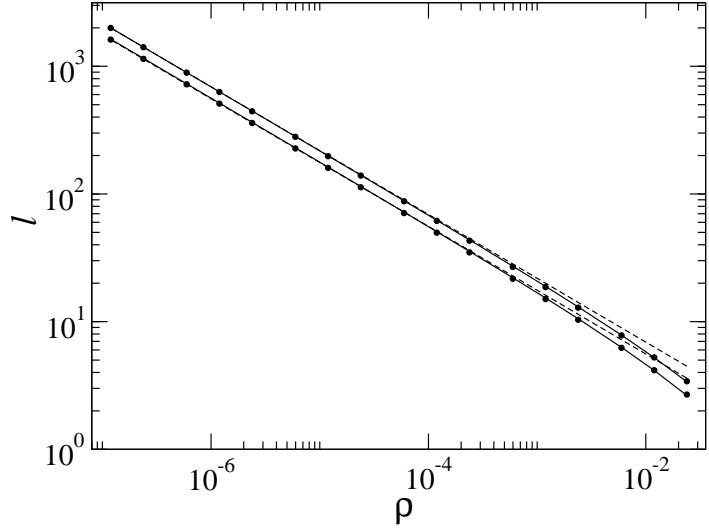}
\caption{\label{fig3d} Plot $\langle |{\bf Y}|\rangle$ (bottom
full circles) and $l=\sqrt{\langle{\bf Y}^2\rangle}$ (top full
circles) obtained from numerical simulations in $D=3$. We compare
these data to the functional forms of Eqs.~(\ref{y13d},\ref{y23d})
(straight dotted lines), finding a good agreement at small
density. Note that a very good fit can be obtained in the entire
range of density by using the functional form
$l_{fit}=l_{theory}-l_0$, where $l_0\approx 1$, for $a=1$ (full
lines).}
\end{figure}
In Fig.~\ref{figpr3d}, we plot the numerical probability density
distribution $p(r)$ for $\rho=10^{-6}$, which compares very well
with our analytical result of Eq.~(\ref{prth}).
\begin{figure}
\psfig{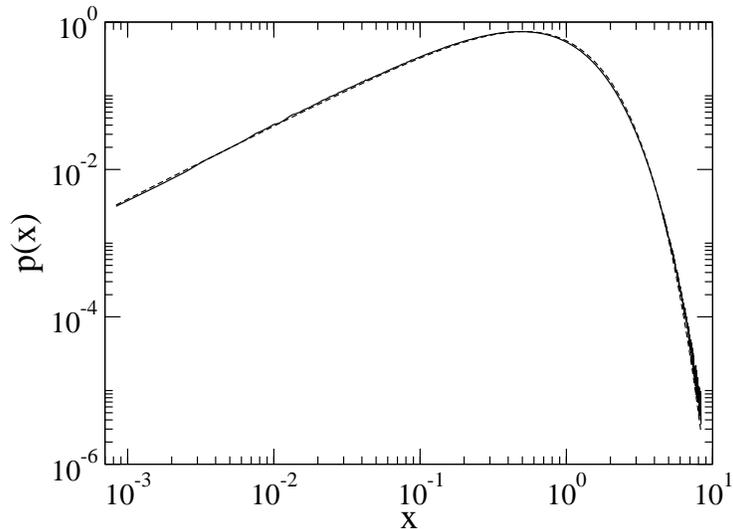}
\caption{\label{figpr3d} In three dimensions and for $\rho=10^{-6}$,
we plot the numerical probability density distribution of the distance
traveled before absorption normalized by its average, $x=Y/\langle
Y\rangle$ (full line), obtained after running a total of $2{\times}
10^8$ random walkers. It is in very good agreement with the
theoretical expression of Eq.~(\ref{prth}), which leads to $p(x)=4x
\exp(-2x)$ (dashed line).}
\end{figure}

\item{$D>3$}: In higher dimensions, we find the same qualitative
behavior as in $D=3$. In particular, the screening length is
\begin{equation}
l= \sqrt{\frac{2D}{\sigma(D)\rho a^{D-2}}},
\end{equation}
and is amplified by a factor $\sqrt{1+\frac{\kappa}{\lambda a}}$
for imperfect absorbers.
\end{itemize}

\section{The case of strongly Correlated absorbers}\label{correla}

\subsection{Exact result in one dimension}

In this section, we consider the case of a non uniform
distribution of absorbers displaying long range density-density
correlations. In one dimension, such a situation arises naturally
when the distribution of intervals $q(L)$ between absorbers decays
as a power law up to a distance $\xi$, which is much larger than
the mean distance between absorbers $\rho^{-1}$. Hence, we take
the typical form
\begin{equation}
q(L)\sim (1-\alpha) \frac{a^{1-\alpha}}{L^{2-\alpha}}\exp(-L/\xi),
\label{distrib1d}
\end{equation}
for $L$ larger than a small scale cut-off $a$ ($a$ should not be
confused with the size of absorbers, which is irrelevant for
perfect absorbers in $D=1$; here we consider point-like
absorbers). We only consider
\begin{equation}
0<\alpha<1.
\end{equation}
The condition $\alpha<1$ ensures that $q(L)$ can be normalized
even in the limit $\xi \to\infty$, whereas $\alpha>0$ implies that
the first moment diverges in this limit:
\begin{equation}
\langle L\rangle =\rho^{-1}=\frac{1-\alpha}{\alpha}
a^{1-\alpha}\xi^\alpha.
\end{equation}
Since $\alpha<1$, $\xi$ is indeed much larger than the mean
distance between absorbers for small $\rho$. If the intervals
between absorbers are drawn independently with the distribution
$q(L)$, the density correlation function $C(r)=\langle \rho({\bf
r})\rho({\bf 0})\rangle$ can be computed exactly, through its
Laplace transform $\hat C(s)$, which is simply related to the
Laplace transform of $q(L)$ by the relation \cite{iia}
\begin{equation}
\hat C(s)=\rho\frac{\hat q(s)}{1-\hat q(s)}.\label{laplace}
\end{equation}
Hence in real space, the connected density correlation behaves
like
\begin{equation}
\langle \rho({\bf r})\rho({\bf 0})\rangle
-\rho^2\sim\rho\frac{a^{\alpha-1}}{r^\alpha}
\exp(-r/\xi)\label{cf1}.
\end{equation}
Now applying the general result of Eq.~(\ref{lexact}), we obtain
\begin{equation}
l=\sqrt{\frac{\alpha}{6(\alpha+2)}}\,\xi\sim
a^{1-1/\alpha}\rho^{-1/\alpha}.
\end{equation}
In one dimension, we find that the screening length scales as the
correlation length, which is much larger than the screening length
obtained in the uniform case. In the next subsections, we shall
illustrate the fact that this very same result should apply in
higher dimensions.

\subsection{A heuristic argument in higher dimensions}\label{gencor}
We consider a system where the absorbers are distributed via a
physical process which leads to long-range spatial correlations
between them. Typical examples are given by directed percolation
(see next subsection), percolation or DLA, the latter being
briefly discussed in section \ref{DLA}. In general, the presence
of correlations makes the problem much more difficult to treat
analytically. Here, we present a semi-phenomenological approach
based on our mean-field renormalization method, which applies in
the case where correlations manifest themselves as a clustering
phenomena. We define the spatial correlation function as
\begin{equation}
\langle \rho({\bf r})\rho({\bf 0})\rangle = \rho c({\bf r}) +
\rho^2, \label{cf}
\end{equation}
where the normalized connected correlation function $c({\bf r})$
is assumed to behave qualitatively as
\begin{equation}
c({\bf r}) \sim{\exp(-r/\xi)\over r^\alpha},\label{cfp}
\end{equation}
where $\xi$ is the correlation length of the system. We also
assume that the system is isotropic. A relationship between the
density and $\xi$ is found by associating $\xi$ as the
characteristic distance where the connected part of the above
correlation function, the first term in Eq.~(\ref{cf}), is of the
same order as the second term. This gives
\begin{equation}
\rho\sim{\xi^{-\alpha }}.  \label{xip}
\end{equation}
We implicitly consider the case where $\xi$ is much bigger than
the mean distance between absorbers, which scales as
$\rho^{-1/D}$. Hence, we will assume from now that
\begin{equation}
\alpha<D.
\end{equation}
The regime of small density thus obviously corresponds to a regime
where the correlation length is large, for instance  near a
continuous transition. If the correlation is manifested by the
formation of clusters, then the typical number of absorbers in a
cluster is given by
\begin{equation}
M_c = \int c({\bf r}) \,d{\bf r} \sim \xi^{D-\alpha},
\end{equation}
which is divergent as $\xi\to \infty$, since $\alpha<D$. If $N_c$
is the number of clusters in the volume ${\cal V}$, the total
number of absorbers is given by $N\sim N_c  \xi^{D-\alpha}$, which
leads to the cluster density
\begin{equation}
\rho_c\sim \rho\,\xi^{\alpha -D}\sim\xi^{-D}. \label{rhoc}
\end{equation}
This last result expresses the fact that the typical distance
between clusters of absorbers is of order $\xi$. We thus expect to
find large empty regions in the system, whose linear size is of
order $\xi$, which is, again, much bigger than the mean distance
between absorbers. In what follows, we will repeat our
renormalization calculation but in terms of the cluster number. As
a starting point, we will use the approximation of
Eq.~(\ref{approx}), where the potential $V$ is concentrated at the
center of the clusters and will take the form
\begin{equation}
V(r) = \lambda {\exp(-r/\xi)\over r^{\alpha}},
\end{equation}
which is simply proportional to the mean density of absorbers for
a cluster whose center is at $r=0$. For large $\lambda$, the flow
equation in $\rho_c$  leads to the same functional form for $g$ as
before, but the corresponding equation for $s= m^2$ is now
\begin{equation}
{\partial m^2\over \partial \rho_c} = {\int c(r) r^{D-1} dr\over
\int c(r)g(r) r^{D-1} dr}. \label{eqtr}
\end{equation}
We first consider the case $\alpha < 2$. In the limit of small $m$
and large $\xi$, we find
\begin{equation}
{\partial m^2\over \partial \rho_c} \sim {\xi^{D-\alpha} (m +
\xi^{-1})^{2-\alpha}}\sim {\xi^{D-2}(1 +m\xi)^{2-\alpha}},
\end{equation}
where we have used $g(r) \sim \exp(-mr)/r^{D-2}$. Using the
relation $\rho_c\sim\xi^{-D}$, we finally obtain
\begin{equation}
m\xi{\partial m\over \partial \xi^{-1}}\sim (1 +m\xi)^{2-\alpha}.
\end{equation}
This homogeneous equation admits the obvious solution
\begin{equation}
m \sim {\xi^{-1}},\label{resxi}
\end{equation}
leading to
\begin{equation}
l\sim\xi\sim a^{1-D/\alpha}\rho^{-1/ \alpha},\label{lxia}
\end{equation}
where we have reintroduced the dependence on the absorber radius
$a$.

In the case where $\alpha>2$, the integral in the denominator of
the right-hand side of Eq.~(\ref{eqtr}) converges and we simply
get
\begin{equation}
m^2 \sim \rho_c \xi^{D-\alpha} \sim\xi^{-\alpha} \sim\rho,
\end{equation}
or
\begin{equation}
l\sim a^{1-D/2}\rho^{-1/2},
\end{equation}
which is the result for uncorrelated absorbers.

For $\alpha=2$ or $D=2$, the calculation above leads to
logarithmic corrections in the expression of $l$. Considering the
crudeness of our argument, we do not believe it is worth detailing
the nature of these corrections.

In conclusion, the present results suggest that the screening
length for a correlated system is either the correlation length
$\xi$ ($\alpha<2$) or the screening length found in the
uncorrelated case ($\alpha>2$). This result can be expressed in a
synthetic way by
\begin{equation}
l\sim \max\left(\xi,a^{1-D/2}\rho^{-1/2}\right),\label{lfinal}
\end{equation}
with logarithmic corrections for the uncorrelated result in $D=2$,
which were obtained analytically in section \ref{sec4}.

\subsection{Numerical results for a critical distribution of
absorbers arising from directed percolation}\label{secperco}

In this section, we will test the ideas presented above by
considering a correlated distribution of absorbers generated by
the active sites remaining at time $t$, at the critical point of
directed percolation \cite{hinri}.

Let us briefly introduce directed percolation on a $D$-dimensional
hypercubic lattice. Lattice sites are empty (inactive) or occupied
by a particle (active). At time $t$, each site is visited in a
parallel dynamics. If the site is occupied, the particle is copied
on its $2D$ neighbors with probability $p$, or removed with
probability $1-p$. If at least one particle has been copied on a
given site, this site is simply considered as occupied, and empty
otherwise. If $p$ is large enough, a finite stationary density of
particles $\rho$ establishes itself for large time. On the
contrary, for small enough $p$, the density decays exponentially
with time. In fact, there exists a critical value $p_c$ for which
the system is critical: the density decreases algebraically, and
the spatial (and temporal) correlation length diverges with time.
This defines the critical exponents $\delta$ and $z$
\begin{equation}
\rho\sim t^{-\delta},\qquad\xi\sim t^{1/z},
\end{equation}
and the density correlation function has the typical form
introduced in Eq.~(\ref{cf},\ref{cfp}), with
\begin{equation}
\alpha=\delta z.
\end{equation}
\begin{figure}
\psfig{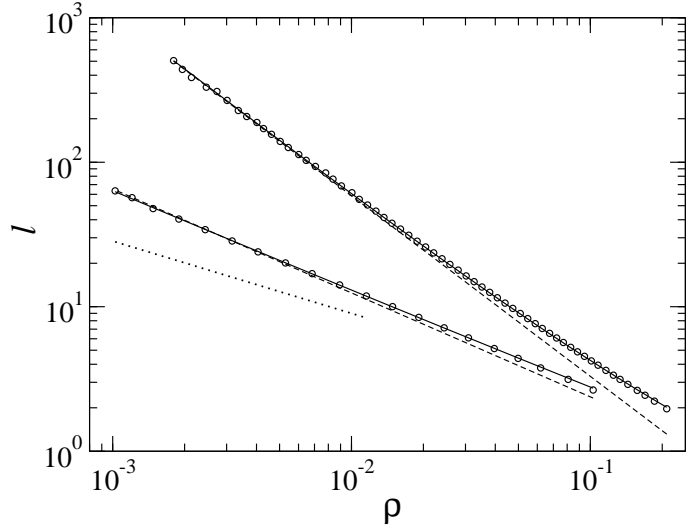}
\caption{\label{figlperco} Plot of $l=\sqrt{\langle{\bf
Y}^2\rangle}$ obtained from numerical simulations of directed
percolation in $D=2$ (top dots; 30 samples on a $4000^2$ lattice)
and $D=3$ (bottom dots; 70 samples on a $300^3$ lattice). Each
point corresponds to an average over a total of $10^7-10^8$ random
walkers, and error bars are smaller than the size of the dots,
except for the smallest densities in $D=2$, where both are of the
same order. The dashed lines have the expected slopes
$1/\alpha=(\delta z)^{-1}$. In both case, the initial curvature
can be well captured by a two-parameter fit to the functional form
$l=A\rho^{-1/\alpha}+B\rho^{-1/2}$, where the second term
corresponds to the uncorrelated screening length (full lines).
Note that the stronger curvature in $D=2$ could as well be due to
subleading logarithmic corrections as suggested in subsection
\ref{gencor}. The dotted line represents the slope $-1/2$, which
is the expected result in the uncorrelated case, and which holds
for $D\geq 4$.}
\end{figure}

We have performed extensive numerical simulations of directed
percolation at the critical point and considered the active sites
present at several fixed times. When any of these times is
reached, we stop the simulation and launch a large number of
random walkers which are absorbed by the active sites. We then
measure $\langle |{\bf Y}|\rangle$ and the screening length
$l=\sqrt{\langle{\bf Y}^2\rangle}$, which are found to be
proportional. Finally, the directed percolation dynamics is
resumed until the next sampling time is reached, permitting us to
explore systems with smaller and smaller densities, and increasing
correlation length. We have performed our simulations in $D=2$ and
$D=3$, since our $D=1$ result being exact, there is no doubt that
the relation $l\sim\xi$ should be satisfied in this case. For
$D<4$ ($D_c=4$ is the critical dimension above which mean-field
theory becomes exact), we have $\alpha<2$. Interestingly, the
equality $\alpha=2$ holds exactly in $D=4$ and above ($\delta=1$
and $z=2$), so that for $D\geq 4$, we recover the result $l\sim
\rho^{-1/2}$, identical to the uncorrelated case. This is
comforting, as correlations are known to become irrelevant above
the upper critical dimension. For the dimensions of interest here,
one has \cite{hinri}
\begin{equation}
\alpha_{D=2}\approx 0.79,\qquad \alpha_{D=3}\approx 1.39.
\end{equation}
In Fig.~\ref{figlperco}, we plot $l=\sqrt{\langle{\bf
Y}^2\rangle}\sim \langle |{\bf Y}|\rangle$ as a function of
$\rho$, and find a fair agreement with our prediction
$l\sim\rho^{-1/\alpha}$. The numerical data are consistent with a
subleading correction of order $\rho^{-1/2}$ (the uncorrelated
result), although the rather strong curvature observed in $D=2$
could be as well ascribed to subleading logarithmic corrections
mentioned at the end of subsection \ref{gencor}.

\begin{figure}
\psfig{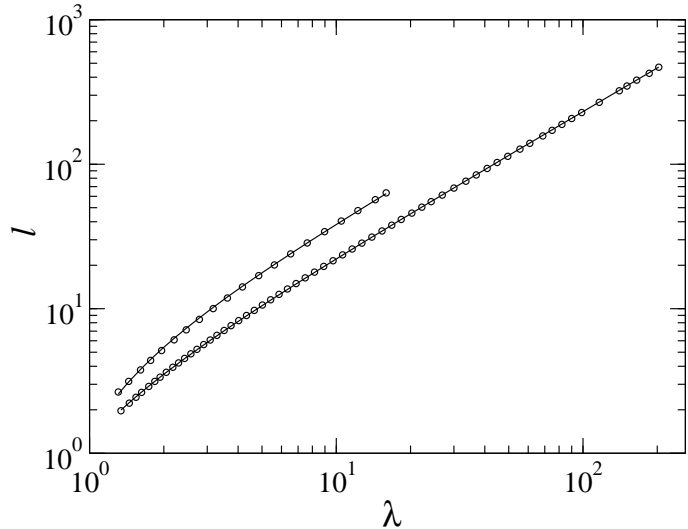}
\caption{\label{figlxiperco} Plot of $l=\sqrt{\langle{\bf
Y}^2\rangle}$ obtained from numerical simulations of directed
percolation in $D=2$ (bottom dots) and $D=3$ (top dots) as a
function of the average void size $\lambda$ (see text), which is
expected to be an alternative measure of $\xi$. The fits to the
linear functional form $l=A\lambda+B$ are excellent (full lines).
The curvature is due to our choice of a log-log plot, in order to
show that the linear fit works well in the entire range of
$\lambda$. Note that $\lambda$ also seems to share the same
statistical fluctuations as $\xi$, so that $\xi$ as a function of
$\lambda$ in $D=2$ is much smoother than when expressed as a
function of $\rho$ (compare with Fig.~\ref{figlperco}, for very
small $\rho$).}
\end{figure}

We have also measured the average void size $\lambda$. We pick a
point at random in space and determine the radius of the largest
disk (in $D=2$) or sphere (in $D=3$) which does not contain any
absorber. In $D=1$ (see Eq.~(\ref{lexactmom1})), this is exactly a
measure of $\xi$ since
\begin{equation}
\lambda= \frac{1}{4}\frac{\langle L^2\rangle}{\langle
L\rangle}=\frac{3}{4}\langle|Y|\rangle\sim l\sim\xi.\label{voids}
\end{equation}
In higher dimension, we expect this property to hold, since
$\lambda$ measures the typical distance between absorber clusters,
as discussed in subsection \ref{gencor}, which was found to be of
order $\xi$. In fact, we postulate the more general result
\begin{equation}
\lambda\sim \max\left(\xi,\rho^{-1/D}\right).
\end{equation}
Fig.~\ref{figlxiperco} illustrates the very convincing linear
relation found numerically between $l$ and $\lambda$, implying
$l\sim\xi$. Note that we have the obvious bound
\begin{equation}
l>\lambda,\label{bound}
\end{equation}
as a particle cannot be absorbed in empty regions.

\subsection{Exact results for some tubular
structures in $D\geq 2$}\label{tubsec}

In this section, we consider a specific geometry in $D\geq 2$
where the screening length can be computed exactly, confirming our
general result of Eq.~(\ref{lfinal}).

Let us start by describing the model in $D=2$. We consider one
dimensional semi-infinite half-lines ($x>0$) of absorbers
separated by a distance $L$ drawn independently with the
distribution $q(L)$. Random walkers start from $x<0$, and the
screening length $l$ is defined as the average depth reached by
the walkers before being absorbed. In this context, $l$ can be
also viewed as a penetration length, similar to the one defined
for DLA (see section \ref{DLA}). A random walker penetrating a
channel of width $L$ will be absorbed at a depth of order $L$, the
only relevant length scale. Actually, the random walker density
$u$ in the tube can be computed exactly, by solving the system
\begin{equation}
\nabla^2 u=0,
\end{equation}
with the absorbing condition for $x>0$
\begin{equation}
u(x,y=0)=u(x,y=L)=0,
\end{equation}
and a constant input of walkers at the entrance of the tube
\begin{equation}
u(x=0,0\leq y\leq L)=1.
\end{equation}
Using discrete Fourier transform along the direction $y$, we
arrive at the result
\begin{equation}
u(x,y)=\frac{4}{\pi}\sum_{n=0}^{+\infty}
\frac{1}{2n+1}{\sin\left(\frac{(2n+1)\pi
y}{L}\right)}\exp\left(-\frac{(2n+1)\pi x}{L}\right),
\end{equation}
which can also be written as a cumbersome expression involving
standard $\ln$ and $\arctan$ functions. The flux of particles
deposited at the depth $x$ is
\begin{equation}
\phi(x)=2{\partial u\over \partial
y}(x,0)=\frac{4}{L\sinh\left(\frac{\pi x}{L}\right)},
\end{equation}
which decays exponentially over the scale $L$. The average
deposition depth is then
\begin{equation}
l(L)=\int_{0}^{+\infty}\frac{4x\,dx}{L\sinh\left(\frac{\pi
x}{L}\right)}=L.
\end{equation}

If we have an array of such channels, the probability that a
walker first enters a channel of width $L$ is
\begin{equation}
Q(L)={L\over \langle L\rangle}q(L).
\end{equation}
If we neglect the process where a random walker leaves this first
channel to be absorbed in an other one, we find
\begin{equation}
l=\int_{0}^{+\infty}Q(L)l(L)\,dL=\frac{\langle L^2\rangle}{\langle
L\rangle},\label{ltubu}
\end{equation}
where $\langle L\rangle$ is the inverse of the absorber density
$\rho$. If $q(L)$ decreases rapidly, $\langle L^2\rangle\sim
\rho^{-2}$, and we find
\begin{equation}
l\sim \rho^{-1}.\label{tubular}
\end{equation}
In this geometry, the absorbers are highly correlated since they
accumulate on lines. Note that the correlation function averaged
over angle initially decreases as $r^{-1}$, so that
Eq.~(\ref{tubular}) is fully consistent with our general result
for $\alpha=1$. Finally, if the width distribution  has a power
law decay as in Eq.~(\ref{distrib1d}), Eq.~(\ref{ltubu}) leads to
\begin{equation}
l\sim \xi \sim\rho^{-1/\alpha},\label{finalrez}
\end{equation}
again in perfect agreement with our general result.

\begin{figure}
\psfig{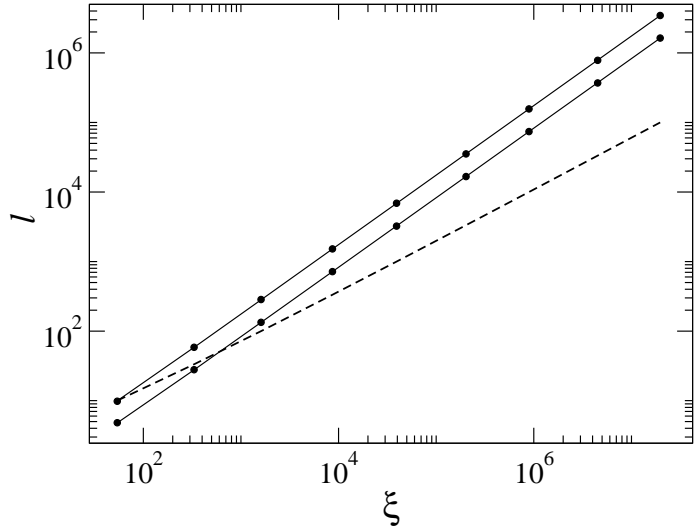}
\caption{\label{figtub} We consider a semi-infinite tubular system
in $D=2$ (see text), where the tube width distribution is
$q(L)\sim L^{\alpha-2}$, for $L\in [1,\xi]$. For $\alpha=3/4$, we
plot $\langle |{\bf Y}|\rangle$ (bottom dots) and
$l=\sqrt{\langle{\bf Y}^2\rangle}$ (top dots) as a function of the
correlation length $\xi$.  The fits to the linear functional form
$l=A\xi+B$ are excellent (full lines). We also show $\langle
L\rangle=\rho^{-1}$ (dashed line), which behaves as $\xi^\alpha$,
for large $\xi$. Each points is an average over at least $10^6$
random walkers, involving a total of $10^7$ absorbing tubes, so
that the error bars are much smaller than the size of the points.}
\end{figure}

In Fig.~\ref{figtub}, we present numerical simulations of this
two-dimensional tubular system. The data are in perfect agreement
with Eq.~(\ref{finalrez}), showing that the processes involving
particles leaving a tube to be absorbed in an other one do not
affect our general result.

\begin{figure}
\psfig{figure=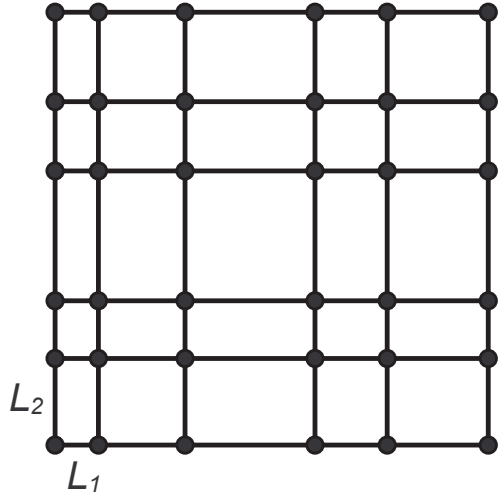,angle=0,height=7cm}
\caption{\label{figprod}In the tubular model introduced in section
\ref{tubsec}, the structure consists of linear tubes whose
$(D-1)$-dimensional section is a cross product generated by $D-1$
linear chains. The absorbers lives on the $(D-2)$-dimensional surface
of ``rectangles'' of size $L_1{\times} L_2{\times} ...{\times}
L_{D-1}$. The figure shows a view from the top of the structure in
$D=3$, where the sections of the tubes are true rectangles. In the
model of section \ref{prodsec}, the $D$-dimensional structure is
generated by placing the absorbers at the intersection of the product
of $D$ linear chains. The figure shows the absorbers (dots) in $D=2$.}
\end{figure}

This model can be generalized in higher dimensions in the
following way. We consider linear tubes whose $(D-1)$-dimensional
surface is perfectly absorbing. A $(D-1)$-dimensional cut of the
system has the structure of a network of $(D-1)$-dimensional
``rectangular'' cells of edge length $L_1$, $L_2$,..., $L_{D-1}$,
which are drawn independently using the same probability
distribution $q(L)$. The absorbers are placed on the
$(D-2)$-dimensional surface of these cells (see
Fig.~\ref{figprod}). The density of absorbers is still
\begin{equation}
\rho=\langle L\rangle^{-1}.
\end{equation}
The solution of the diffusion problem in each tube is simply
\begin{equation}
u(x,y_1,...,y_{D-1})=\left(\frac{4}{\pi}\right)^{D-1}
\sum_{n_1...n_{D-1}=0}^{+\infty}
\prod_{j=1}^{D-1}\frac{1}{2n_j+1}\sin\left(\frac{(2n_j+1)\pi
y_j}{L_j}\right)\exp\left(-\pi x\sqrt{\sum_{k=1}^{D-1}
\frac{(2n_k+1)^2}{L_k^2}}\right),
\end{equation}
which decays exponentially on the scale $l({\bf L})$, with
\begin{equation}
l({\bf L})=\min_{j=1,..., D-1} L_j
\end{equation}
Hence,  neglecting again processes where a walker leaves the first
tube it enters to be absorbed elsewhere, the average penetration
length is
\begin{equation}
l=\langle L\rangle^{-(D-1)}
\int_{0}^{+\infty}dL_{min}\,L_{min}^2q(L_{min})
\left[\int_{L_{min}}^{+\infty}dL\,Lq(L)\right]^{D-2}.\label{newl0}
\end{equation}
If $q(L)$ decays rapidly for $L\gg \langle L\rangle$, this
integral leads to
\begin{equation}
l\sim \langle L\rangle\sim\rho^{-1}.
\end{equation}
However, if $q(L)$ takes the form of Eq.~(\ref{distrib1d}), we
find
\begin{eqnarray}
l&\sim &\langle L\rangle^{-(D-1)}
\int_{0}^{+\infty}dL\,L^2L^{\alpha-2}\exp(-L/\xi)
\left[L^{\alpha}\exp(-L/\xi)\right]^{D-2},\\
&\sim & \langle L\rangle^{-(D-1)}\xi^{(D-1)\alpha+1}\sim\xi,
\end{eqnarray}
where we have used the fact that $\langle
L\rangle\sim\xi^\alpha\sim\rho^{-1}$. We again find $l\sim \xi$,
in agreement with our general argument for strongly correlated
absorbers.

\subsection{Exact results for a product
distribution of absorbers in $D\geq 2$}\label{prodsec}

In this section, we consider a system for which absorbers are
placed at the intersections of the product of $D$ linear chains
(see Fig.~\ref{figprod}). Each chain has its intervals drawn from
the distribution $q(L)$. The density of absorbers reads
\begin{equation}
\rho=\langle L\rangle^{-D}.
\end{equation}
If the distribution $q(L)$ decreases rapidly on the scale of
$\langle L\rangle$, it is clear that this system will behave
similarly to a uniform distribution of absorbers, leading to
$l\sim\rho^{-1/2}$. Hence, we assume that the distribution of
intervals is of the form
\begin{equation}
q(L)\sim {L^{-2+\alpha/D}}\exp(-L/\xi). \label{distribd}
\end{equation}
We shall see below that $\alpha<2$ will lead to $l\sim \xi$,
whereas the uncorrelated result is recovered for $\alpha>2$. The
correlation function can be exactly computed, since the density is
the cross product of $D$ independent densities
\begin{equation}
\langle \rho({\bf x})\rho({\bf 0})\rangle =\prod_{i=1}^D
C(x_i),\qquad {\bf x}=(x_1,...,x_D),
\end{equation}
where $C(x)$ is the one dimensional correlation function, which
can be exactly computed in terms of the Laplace transform of
$q(L)$ (see Eq.~(\ref{laplace})). After performing an average over
angle, the correlation is found to qualitatively behave as
\begin{equation}
\langle \rho({\bf r})\rho({\bf 0})\rangle
-\rho^2\sim\frac{\rho}{r^\alpha} \exp(-r/\xi)\label{cf3}.
\end{equation}
In the present model, the average void size introduced in section
\ref{secperco} is defined as the radius of the largest hypercube that
does not contain any absorbers, averaged over the position of the
center of the hypercube. A simple calculation leads to the
generalization of Eq.~(\ref{voids}). If the center ${\bf
x}=(x_1,...,x_D)$ ($x_i\in [0;L_i]$) is drawn randomly in a cell of
size $L_1{\times} L_2{\times} ...{\times} L_{D}$ with all $2^D$
corners occupied by an absorber, the largest hypercube not containing
any absorber has a radius
\begin{equation}
d({\bf x})=\min_{k=1,..., D}\min(x_k,L_k-x_k).
\end{equation}
In order to simplify our calculation, we replace the actual
expression of $d({\bf x})$, by
\begin{equation}
d({\bf x})=\min(x_k,L_k-x_k),\qquad L_k=\min_{j=1,..., D} L_j,
\end{equation}
which behaves essentially in the same manner as the original
expression. Since the average of $d({\bf x})$ over the position
${\bf x}$ of the center is $L_k/4$, we finally get
\begin{equation}
\lambda=\frac{1}{4}\langle L\rangle^{-D}
\int_{0}^{+\infty}dL_{min}\,L_{min}^2q(L_{min})
\left[\int_{L_{min}}^{+\infty}dL\,Lq(L)\right]^{D-1}.\label{newl}
\end{equation}
Interestingly, Eq.~(\ref{newl}) is the same (up to the $1/4$
factor) as the expression obtained for $l$ in the previous
subsection (see Eq.~(\ref{newl0})), except that $D-1$ is now
replaced by $D$. Hence, if $\alpha<2$, we conclude that
\begin{equation}
l>\lambda\sim\xi\sim\rho^{-1/\alpha}.
\end{equation}
In fact, our previous results suggest that $l\sim\lambda\sim\xi$
again. Finally, note that in an homogeneous correlated system,
Eq.~(\ref{newl}) shows that the distribution of the void linear
sizes is typically of the form
\begin{equation}
Q(L)\sim \frac{L^{\alpha-1}}{\xi^\alpha}\exp(-L/\xi),\label{voiddist}
\end{equation}
which is identical to the void size distribution $Q(L)$ obtained in
$D=1$ (see Eq.~(\ref{qq})).

\begin{figure}
\psfig{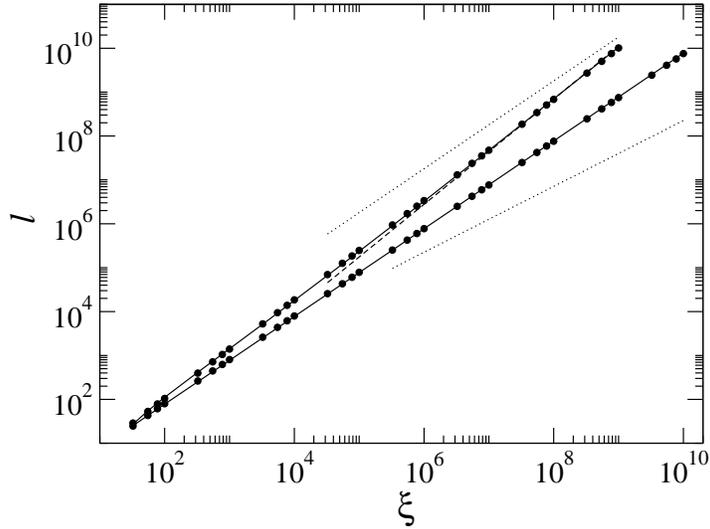}
\caption{\label{figlprod} Plot of $l=\sqrt{\langle{\bf
Y}^2\rangle}$ as a function of $\xi$ obtained from numerical
simulations (full circles) for the model where absorbers are
placed at the intersections of the cross product of $D$ linear
chains in $D=3$. For $\alpha=2.4>2$ (top circles and curves), we
find an asymptotic slope $l\sim \xi^{\alpha/2}\sim \rho^{-1/2}$
(top dashed line). The fit to the functional form
$l=A\xi^{\alpha/2}+B\xi$ (top full line; where the linear
correction corresponds to the correlated result) is excellent. As
a guide to the eye, we show the line of slope unity corresponding
to the correlated regime $l\sim\xi$ (top dotted line). For
$\alpha=1.5<2$ (bottom circles and curves), we find that $l=A\xi$
fits perfectly the data (bottom full line). We also plot a line of
slope $\alpha/2=0.75$ (bottom dotted line) to illustrate that
$l\sim\xi$ grows much faster than the uncorrelated screening
length $\rho^{-1/2}$.}
\end{figure}

Note that in $D>2$, we can obtain a regime where the correlation
length is large compared to the average distance between absorbers,
but is still smaller than the uncorrelated screening length,
\begin{equation}
\rho^{-1/D}\ll\xi\ll\rho^{-1/2}.
\end{equation}
This regime corresponds to values of $\alpha$ satisfying,
\begin{equation}
2<\alpha<D.
\end{equation}
In this case, although there are long-range correlations in the
system, our argument of section \ref{gencor} predicts that the
uncorrelated result $l\sim \rho^{-1/2}\sim \xi^{\alpha/2}$ should
hold.  In Fig.~\ref{figlprod}, we measure numerically $l$ as a
function of $\xi$ for the model studied in the present section,
and in $D=3$. In perfect agreement with our general result of
Eq.~(\ref{lfinal}), we find that $l\sim \xi$ for $\alpha=1.5<2$,
whereas $l\sim \rho^{-1/2}$ for $\alpha=2.4>2$. Note that in both
cases, we find $\lambda\sim\xi\gg\rho^{-1/D}$ (not shown).

\subsection{Temporal decay of the density of random
walkers for a correlated density of absorbers}

In this subsection, we address the generalization of
Eq.~(\ref{don}) for a strongly correlated distribution of
absorbers in $D$ dimensions. We shall adapt to the present problem
the usual variational argument \cite{donsker} leading to
Eq.~(\ref{don}). Note that the effect of short range clustering on
the density of surviving random walkers has been studied
quantitatively in \cite{new}.

Imagine that one releases a random walker of diffusion constant
$\kappa$ in an hypercubic region of linear size $L$ not containing any
absorber (a void). Its survival probability is larger than the
survival probability computed assuming that the boundary of the void
is perfectly absorbing. This probability behaves like $P(t)\sim
\exp(-\kappa {\bf k}_L^2t)$, where ${\bf k}_L$ is the lowest mode of
the diffusion equation in the hypercubic domain delimited by the void,
with absorbing boundary conditions on the surface. All the $D$
coordinates of ${\bf k}_L$ are equal to $\pi L^{-1}$. By actually
solving the diffusion equation mentioned above with a uniform initial
position of the random walker, we obtain an exact bound of the form
\begin{equation}
P(t)> \frac{A_D}{L^D}\exp\left(-\frac{\pi^2\kappa Dt}{L^2}\right),
\end{equation}
where $A_D$ is some constant which does not depend on $L$

Finally, and after averaging over the void size distribution
$Q(L)$, we find an exact bound for the density of surviving
walkers
\begin{equation}
n(t)>A_D\int_{0}^{+\infty}Q(L) \exp\left(-\pi^2\frac{\kappa
Dt}{L^2}\right)\,dL.
\end{equation}
It is commonly conjectured that this kind of bound actually
captures the correct asymptotic behavior of $n(t)$ \cite{donsker}.

Let us now assume a distribution of hypercubic void sizes of the form
found in Eq.~(\ref{voiddist}).
\begin{equation}
Q(L)\sim
\frac{L^{\alpha-1}}{\xi^\alpha}F(L/\xi).\label{voiddistnew}
\end{equation}
In $D=1$, or in the preceding section, we considered a cut-off function
$F(x)=\exp(-x)$. Here, we will consider the more general case
\begin{equation}
F(x)=\exp(-x^{\gamma}),
\end{equation}
where the value $\gamma=D$ is a natural example: $L^D/\xi^D$ is
the volume of the void divided by the correlation volume, an
equivalent of the term $\rho L^D$ obtained in the usual
uncorrelated case (see Eq.~(\ref{volimp})).

Dropping all unimportant numerical constants for the sake of clarity,
we obtain
\begin{equation}
n(t)\sim \int_{0}^{+\infty}L^{-D-1+\alpha}\exp\left(-\frac{\kappa
t}{L^2}-\left(\frac{L}{\xi}\right)^\gamma\right)\,dL.
\end{equation}
Applying a saddle-point argument, and dropping subleading
corrections, we finally find that
\begin{equation}
n(t)\sim \exp\left(-C_D\left(\frac{\kappa
t}{\xi^2}\right)^{\frac{\gamma}{\gamma+2}}\right).
\end{equation}
We obtain a stretched exponential decay like in the uncorrelated
case, but more importantly, we find that the time scale $\tau$
over which the density decays is now controlled by the correlation
length
\begin{equation}
\tau=\frac{\xi^2}{\kappa},
\end{equation}
instead of the density $\rho$, as obtained in the uncorrelated case
\begin{equation}
\tau=\frac{\rho^{-2/D}}{\kappa}.
\end{equation}

\section{Penetration length for DLA}\label{DLA}

There have been several attempts to measure the screening length
of DLA clusters \cite{wit} of gyration radius $R$
\cite{lintro,meakin,llast,bayard,locgrowth}. In this section, we
show that the methods used so far {\it do not} effectively measure
$l$. We will propose a theoretical estimate for $l$ as well as a
possible numerical method in order to measure it properly.

First of all, let us briefly mention how DLA clusters are grown in
dimension $D\geq 2$. One first places a seed at the origin, for
instance a $D$-dimensional sphere of radius $a$. A random walker
(of same radius $a$) is launched from far away and wanders until
it touches the seed, whereupon it sticks to it. Then, another
random walker is released, which sticks upon contact with any of
the quenched particles. This process goes on, leading to the
formation of fractal clusters of dimension $D_f<D$ (and
presumably, $D_f>D-1$ \cite{wit}), defined as
\begin{equation}
N\sim R^{D_f},
\end{equation}
where $N$ is the number of particles in the cluster. We
immediately note that for an incoming random walker, the points at
a distance $2a$ from the already formed structure acts like
absorbers. Hence, it is tempting to introduce a screening length
$l$ \cite{lintro,meakin}, measuring how deep the random walkers
penetrate the cluster before sticking to it. However, contrary to
the systems studied so far in this paper, which were infinite and
homogeneous, a DLA cluster is finite and its density from the
center decays in average as
\begin{equation}
\rho(r)\sim r^{-(D-D_f)}.\label{rhodla}
\end{equation}
The determination of $l$ is an important matter, being often an
essential ingredient in the theoretical attempts to determine
$D_f$. Although mean-field theories are based on contradicting
estimates for $l$, typically $l\sim R^{D-D_f}$ \cite{muka,mf1} or
$l\sim R^{(D-D_f)/2}$ \cite{mf2}, they mostly consider that $l\ll
R$. However, there is now convincing numerical evidence
\cite{meakin,llast,bayard} showing that
\begin{equation}
l\sim R,
\end{equation}
at least in $D=2$, and probably in $D=3$. To understand this
result, it is worth mentioning the commonly used method to
determine $l$ \cite{meakin,llast}. A cluster of large size $R$ is
first grown. Then a large number of independent random walkers are
released. If one of them touches a particle of the cluster, it
simply vanishes, and its last position is recorded. The
distribution of the positions of absorption is thus obtained and
its standard deviation is considered as a measure of $l$ (before
or after averaging over many clusters having the same number of
particles). Although a certain consensus seems to exist in the
literature, we do not believe that this variance is a faithful
measure of $l$. To see this, let us consider the simple example of
a perfectly absorbing ellipse in $D=2$ of aspect ratio $e>1$ and
main axis $R$. Any method to measure $l$ should retrieve the
trivial result that $l=0$ in this case.  However, the method
presented above would obviously lead to $l\sim R\,$! It is correct
that DLA clusters grown in the continuum are statistically
isotropic. However, for a given cluster, the fluctuation of the
length of the main branches are typically of size $R$, leading to
an automatic numerical evaluation of $l\sim R$. In other words,
even if the random walkers were only absorbed near the tip of
these branches, physically implying a very small value of $l$, the
current method of estimating $l$ would invariably lead to $l\sim
R$.

As an alternative method (but which might prove as ineffective as the
one above), we propose that one should measure the distance of
penetration from the {\it convex hull} of the cluster, and to measure
the variance of this length {\it before} averaging over many clusters.
At least, this method gives the correct result for the ellipse (its
own convex hull), that is, $l=0$.

Finally, inspired by the general conclusions of the previous
sections, we would like to give three related arguments in favor
of the result $l\sim R$, which we think is correct, although it is
argued that the available numerical simulations using the method
exposed above are not conclusive on this matter. Particles in a
DLA cluster are obviously strongly correlated, arranging
themselves in highly ramified structures. It is thus tempting to
apply some of our result to this situation:

\begin{itemize}
\item There is no finite correlation length in a DLA cluster
except, rather trivially, for the size of the cluster itself. For
long-range correlations, we have found that $l\sim\xi\sim R$.

\item In a DLA cluster, the density correlation function decays as
$c(r)\sim r^{-\alpha}$, with $\alpha=D-D_f$. Our study suggests
that $l\sim \rho^{-1/\alpha}\sim R$ (see Eq.~(\ref{rhodla})).

\item Since a DLA cluster is fractal, the largest voids inside it
are of typical linear size $\lambda \sim R$. In homogeneous
structures, $l>\lambda$, which suggests again that for DLA, $l
\sim R$.

\end{itemize}

\section{Conclusion}

In this paper, we have justified analytically the heuristic
argument presented in the introduction to estimate the screening
length in a system of uniformly distributed perfect or imperfect
absorbers. Our results in $D=2$ and $D=3$ are in excellent
agreement with numerical simulations. Even the numerical
prefactors of the two first moments are surprisingly well
described by our approach. As a further check, we found that the
theoretical distribution of the distances of absorption is in
perfect agreement with numerical simulations.

For correlated absorbers, our analytical approach is not as
rigorous as in the uncorrelated case, although we supplemented our
heuristic argument with the exact solution in $D=1$ as well as for
two toy models in $D>1$, which fully confirm our general results.
We find that if the density correlation function decays with an
exponent $\alpha$ ($\alpha<2$ in $D\geq 2$, and $\alpha<1$ in
$D=1$) up to the correlation length $\xi\sim \rho^{-1/\alpha}$,
then the screening length scales as $l\sim\xi\sim\lambda$, where
$\lambda$ is the average void linear size. These results were
confirmed numerically in directed percolation, where the active
sites play the role of the absorbers.

Finally, we have argued that for DLA, the penetration or screening
length $l$ should be of the same order as the linear size $R$ of a
cluster. We have also emphasized that the current numerical method
of determining $l$, although finding just $l\sim R$, could hardly
lead to any other result. As a more appropriate method, we propose
measuring the distance of penetration from the convex hull of the
cluster, and to define $l$ as the standard deviation of this
length before averaging over many clusters. With this definition,
the behavior of $l$ as a function of the cluster radius $R$ is
certainly a question worth investigating \cite{unpub}.

\acknowledgments

We would like to thank one Referee whose remarks lead us to also
consider the case of imperfect absorbers ($\lambda<+\infty$).


\begin{thebibliography}{0}


\bibitem{donsker} B.~Ya. Balagurov and V.~G. Vaks,
{Zh. Eksp. Teor. Fiz.}  {\bf 65}, 1939 (1973) [{Sov. Phys. JETP}
{\bf 38}, 968 (1974)]; M. Donsker and S. Varadhan, {Commun. Pure
Appl. Math.} {\bf 28}, 525 (1975); P. Grassberger and I.
Procaccia, {J. Chem. Phys.} {\bf 77}, 6281 (1982); T.~C.
Lubensky, {Phys. Rev. A} {\bf 30}, 2657 (1984); S. Renn, {Nucl.
Phys.  B} {\bf 275}, 273 (1986); J.~W. Haus and K.~W. Kehr, {
Phys. Rep.} {\bf 150}, 263 (1987); T.~M. Nieuwenhuizen, {Phys.
Rev. Lett.} {\bf 62}, 357 (1989).

\bibitem{muka} M. Muthukumar, Phys. Rev. Lett. {\bf
50}, 839 (1983).

\bibitem{dean} D.~S. Dean, I.~T. Drummond, R.~R. Horgan, and
A. Lef\`evre, J. Phys. A {\bf 37}, 10459 (2004).

\bibitem{itzy} C. Itzykson and J.-M. Drouffe, {\it Statistical Field Theory},
Volume 1  (Camdridge, 1992).

\bibitem{feller} W. Feller, {\it An Introduction to Probability Theory
and Its Applications}, Volume 1  (Wiley, New York, 1957).

\bibitem{abrom} M. Abramowitz and I.~A. Stegun, {\it Handbook of mathematical
functions} (Dover, 1965).

\bibitem{iia} S.~N. Majumdar, C. Sire, A.~J. Bray, and S.~J. Cornell,
Phys. Rev. Lett. {\bf 77}, 2867 (1996); B. Derrida, V. Hakim, and
R. Zeitak, Phys. Rev. Lett. {\bf 77}, 2871 (1996).

\bibitem{hinri} H. Hinrichsen, Adv. Phys. {\bf 49}, 815 (2000).

\bibitem{new} Yu.~A. Makhnovskii, A.~M. Berezhkovskii, D.-Y. Yang,
S.-Y. Sheu, and S.~H. Lin,  Phys. Rev. E {\bf 61}, 6302 (2000).

\bibitem{wit} T.~A. Witten and L.~M. Sander, Phys. Rev. Lett. {\bf
47}, 1400 (1981).

\bibitem{lintro} M. Plischke and Z. R\'acz, Phys. Rev. Lett. {\bf
53}, 415 (1984).

\bibitem{meakin} S. Tolman and P. Meakin, Phys. Rev. A {\bf 40}, 428 (1989).

\bibitem{llast} R.~C. Ball, N.~E. Bowler, L.~M. Sander, and E. Somfai,
Phys. Rev. E {\bf 66}, 026109 (2002).

\bibitem{bayard} B.~K. Johnson, R.~F. Sekerka, and M.~P. Foley,
Phys. Rev. E {\bf 52}, 796 (1995).

\bibitem{locgrowth} J. Lee, S. Schwarzer, A. Coniglio, and H.~E. Stanley,
Phys. Rev. E {\bf 48}, 1305 (1993).

\bibitem{mf1} M. Matsushita, K. Honda, H. Toyoki, Y. Hayakawa, and
H. Kondo, J. Phys. Soc. Japan {\bf 55}, 2618 (1988).

\bibitem{mf2} H.~G.~E. Hentschel, Phys. Rev. Lett. {\bf 52}, 212
  (1984); M. Tokuyama and K. Kawasaki, Phys. Lett. A {\bf 100}, 337
  (1984); K. Ohno, K. Kikuchi, and H. Yasuhara, Phys. Rev. A {\bf 46},
  3400 (1992).

\bibitem{unpub} J. Sopik, C. Sire, and D.~S. Dean, in preparation.

\end{thebibliography}
\end{document}